\documentclass[prl,aps,twocolumn,groupaddress]{revtex4-1}
\usepackage{latexsym}
\usepackage{graphicx}
\usepackage{verbatim}
\usepackage{multirow}
\usepackage{amsmath}
\newcommand{\beq}{\begin{eqnarray}}
\newcommand{\eeq}{\end{eqnarray}}
\usepackage{mathrsfs}
\usepackage{float}
\usepackage[usenames, dvipsnames]{color}
\usepackage{mathtools}
\usepackage{slashed}
\usepackage{physics}	% installed packages for typing maths and physics
\usepackage{graphicx}   % need for figures
\usepackage{epstopdf}
\usepackage{subfigure}  % use for side-by-side figures
\usepackage{hyperref}   % use for hypertext links, including those to external documents and URLs
\usepackage{bbold}
\usepackage{wasysym}

\begin{document}

\title{Absence of  a Charge Diffusion Pole \textcolor{black}{at Finite Energies} in an Exactly Solvable Interacting Flat Band Model in d-dimensions}
\author{Philip W. Phillips}
\affiliation{Department of Physics, University of Illinois at Urbana-Champaign, Urbana, Illinois, USA}
\author{Chandan Setty}
\affiliation{Department of Physics, University of Illinois at Urbana-Champaign, Urbana, Illinois, USA}
\author{Shuyi Zhang}
\affiliation{Department of Physics, University of Illinois at Urbana-Champaign, Urbana, Illinois, USA}

\begin{abstract}
Motivated by recent bounds for charge diffusion in critical matter, we investigate the question:  What sets the scale for charge diffusion in a scale-invariant system?    To make our statements precise, we analyze the diffusion pole in an exactly solvable model for a Mott transition in the presence of a \textcolor{black}{long-range} interaction term. To achieve scale invariance, we limit our discussion to the flat-band regime.  We find in this limit that the diffusion pole which would normally obtain at finite \textcolor{black}{energy} is pushed to zero \textcolor{black}{energy} resulting in a vanishing of the diffusion constant. \textcolor{black}{This occurs even in the presence of interactions in certain limits, indicating the robustness of this result to the inclusion of a scale in the problem.}  Consequently, scale-invariance precludes any reasonable definition of the diffusion constant.  Nonetheless, we do find that a scale can be defined, all be it, irrelevant to diffusion, which is the product of the \textcolor{black}{squared} band velocity and the density of states.  \end{abstract}

\maketitle

Scale invariance is both a simplifying and problematic organizing principle for strongly correlated systems.  On the one hand, it dictates that the correlation functions must obey power-law decay with a universal length scale  but on the other, it precludes the presence of a natural energy scale from entering the transport properties.  For electronic systems, this implies that the Fermi energy cannot enter any transport property if scale invariance is present.  This is particularly problematic in describing the strange metal in the cuprates as both scale-invariance and a breakdown of the particle concept have been advocated\cite{homes2004,qcrit2,Marel2003,Valla2110,Anderson1997,abrahams,pll2015} to be operative.  In fact, the key characteristic of quantum critical systems, namely the presence of a dissipation rate that scales linearly with temperature, the Planckian limit of dissipation, has been shown\cite{zaanen} to undergird the experimental observation of Homes' law\cite{homes2004} in the cuprates.  Since Homes' law is about the dc conductivity just above the normal state, some natural scale should govern charge diffusion in critical matter. Hence the question as to what sets the scale for transport in quantum critical systems and scale-invariant systems emerges more generally.   Since the Planckian\cite{zaanen} rate, $\hbar/k_BT$ is the natural scale for dissipation, a secondary question is does this set an upper bound for dissipation and hence serve to bound any subsequent charge diffusion?   Neither of these questions has been answered definitively since violations\cite{pakhira2014,kuang2015,kovtun2008,kovtun2003,myers2007,pang2009} to charge diffusion bounds\cite{hartnoll,blake}, based on the Planckian upper bound coupled with a diffusion constant parameterized by some phenomenological velocity, abound. 

In this paper, we limit our discussion to the former question, namely what sets the scale in scale-invariant systems for charge diffusion?  To make our conclusions precise, we focus on an exactly solvable model for strongly correlated electrons that exhibits a metal-insulator transition regardless of the dimension.  To impose scale invariance, we focus on the flat-band limit.  In the flat band limit, the energy-spacing between energy levels, $\Delta E$, is the smallest scale in the problem and as a consequence, the band velocity vanishes.
 Hence, the  resultant diffusion constant defined by $D\sim v^2\tau$ must vanish if $\tau$ is finite. The scattering time in our problem is governed by scalar impurity interactions and hence is finite. We show explicitly that even for an interacting system, \textcolor{black}{in certain limits and for certain forms of the interaction}, the diffusion pole which would normally occur at finite \textcolor{black}{energy} is pushed to zero-\textcolor{black}{energy} and continues to be dictated by scale invariance.  Hence, \textcolor{black}{scale invariance is robust to interactions in these limits}, and strictly \textcolor{black}{speaking}, there is no energy scale that emerges which permits a reasonable definition of the charge diffusion constant in \textcolor{black}{such a} scale-invariant system.  Nonetheless, we do find that a flat-band scale can be defined, \textcolor{black}{all be it, irrelevant to diffusion}, if one were to consider the product of the band velocity \textcolor{black}{square} and the density of states. Since the density of states diverges, the product can be finite. \textcolor{black}{This \textit{flat band constant} appears as the residue of a diffusion pole that has been shifted to zero energy.} We analyze \textcolor{black}{the properties of} this constant here and show that \textcolor{black}{it can be effectively enhanced, suppressed or unaffected depending on the nature of the long range interaction.}

The model\cite{hk1992,hk1996} we analyze has long-range non-local interactions  with standard tight-binding hoppings, 
 \begin{eqnarray}\nonumber
	H &=& -t \sum_{\langle j,l\rangle,\sigma} \pqty{ c^\dagger_{j\sigma} c^{}_{l\sigma} + h.c. }  - \mu \sum_{j\sigma} c^\dagger_{j\sigma} c^{}_{j\sigma}\\
&& + \frac{U}{N} \sum_{j_1..j_4}\delta_{j_1+ j_3, j_2+ j_4} c^\dagger_{j_1\uparrow} c^{}_{j_2\uparrow} c^\dagger_{j_3\downarrow} c^{}_{j_4\downarrow},\\ \nonumber
	%=& \sum_{k,\sigma} (\epsilon_k - \mu) c^\dagger_{k,\sigma} c^{}_{k,\sigma} + U \sum_k c^\dagger_{k,\uparrow} c^{}_{k,\uparrow} c^\dagger_{k,\downarrow} c^{}_{k,\downarrow} \\
	%\equiv & \sum_{k,\sigma}\, \xi_k n_{k,\sigma} + U \sum_k n_{k,\uparrow} \, n_{k,\downarrow}	
\end{eqnarray}
where the first and second terms denote the local hopping and chemical potential and are set by the scale $\gamma$. The last term is the infinite-range Hubbard-like interaction $U$; this term is non-zero for electrons that scatter in such a way that their position vectors satisfy the constraint of center of mass conservation given by $j_1+j_3 = j_2 + j_4$.  This model predates the SYK\cite{sy1993,k2015} model by 2 years,
though it is considerably less studied. Although both models contain non-local interactions, the current model is exactly solvable as a result of the conservation of the center of mass in the interaction term. Similar models with long range correlations were studied in Refs \cite{baskaran1991exactly,muthukumar1994toy,continentino1994scaling}. The integrability of this model, without resorting to a $1/N$ expansion as in the SYK model\cite{sy1993,k2015}, is best seen in momentum space
\begin{equation}
	H = \sum_{\vec k}H_{\vec k} = \sum_{\vec k} \left (\xi(\vec k)(\hat{n}_{\vec k\uparrow} + \hat{n}_{\vec k\downarrow} ) + U  \hat{n}_{\vec k\uparrow} \, \hat{n}_{\vec k\downarrow}\right),
\label{eq:kSpaceHK}
\end{equation}
from which it is clear that the kinetic and potential energy terms commute.   Here  $\xi(\vec k) \equiv \epsilon(\vec k ) - \mu$ and $\hat{n}_{\vec k \sigma} \equiv c_{\vec k \sigma}^{\dagger} c_{\vec k \sigma}$.  We see that in this model, different momentum states are decoupled, and the Hamiltonian can be diagonalized by states in the number representation for each $\vec k$. The basis of states which spans Eq.~\ref{eq:kSpaceHK} are given by $\left(\mid0\rangle_{\vec k}, \mid \uparrow\rangle_{\vec k}, \mid\downarrow \rangle_{\vec k}, \mid\uparrow\downarrow \rangle_{\vec k}\right)$ with eigenvalues $\left(0,\xi(\vec k),\xi(\vec k), 2 \xi(\vec k) + U\right)$, for each momentum point $\vec k$.  Regardless of the simplicity of this model, a non-trivial Mott transition exists as can be seen from the single-particle Green function,
\begin{equation}
G_0(\vec k,i\omega_n)_U = \left(\frac{g(\vec k, U)}{i \omega_n - \xi(\vec k)+\frac{U}{2}} +  \frac{1- g(\vec k, U)}{i\omega_n - \left(\xi(\vec k) - \frac{U}{2}\right)}\right),
\label{eq:GreensFunction}
\end{equation}
where the function $g(\vec k, U)$ is defined by
\begin{equation}
g(\vec k, U) = \frac{1+ e^{-\beta \xi(\vec k)}}{1+ 2 e^{-\beta \xi(\vec k)} + e^{-\beta\left(2\xi(\vec k) + U\right)}},
\end{equation}
with $\beta$ being the inverse temperature, and $i\omega_n$, the fermionic Matsubara frequency. At half-filling, the Green function is a sum of poles at $E^\pm_k=\xi(\vec k)\pm U/2$.  A gap exists between the two bands when $U> 4td$, that is, when $U$ exceeds the non-interacting bandwidth.  The transition to the gapped state is of the Mott type because ${\cal \Re }G(\omega,\vec k)=0$ identically when $\xi_{\vec k}=0$.  That is, in the gapped state, the Fermi surface of the non-interacting system is converted into a surface of zeros, the fingerprint\cite{stanescu2007} of Mottness. This coincidence obtains entirely because of particle-hole symmetry\cite{stanescu2007} of the underlying Hamiltonian.  We have as our starting point then an exactly solvable model which exhibits a Mott transition regardless of the spatial dimension.

Before we use this model to analyze the existence or lack there of a diffusion pole, we review the standard formulation for a free-electron gas in which scalar impurities act as the source of momentum relaxation. One natural way to obtain such a response is to extend the corresponding density response function of the electron liquid by replacing the infinitesimally small adiabatic continuation parameter $\eta$ with the inverse impurity scattering life time $\frac{1}{\tau}$, i.e., by making the substitution $\chi^{imp}(\vec q, \omega + i\eta) \rightarrow \chi(\vec q, \omega + \frac{i}{\tau})$, where $\chi(\vec q, \omega)$ ($\chi^{imp}(\vec q, \omega)$) is the density response function of the electron liquid without (with) impurities. However, it was pointed out~\cite{mermin1970lindhard} that such a naive substitution does not respect the continuity equation. %A violation of the continuity equation is definitely not reasonable since, in the presence of impurities, even though the crystal momentum is not a good quantum number, it is imperative that the electron number be locally conserved. From a linear response perspective, there must exist a definite relationship between the longitudinal current response and the density-density response of the form~\cite{giuliani2005quantum} $\chi(\vec q, \omega) = \frac{q^2}{\omega^2} \chi_L(\vec q, \omega)$, where $\chi_L(\vec q, \omega)$ is the longitudinal current correlation function. It can be concluded that such a relationship is guaranteed through gauge invariance by observing that a longitudinal vector potential has the same physical effects as that of a scalar potential, and that a current response induced by such a vector potential also induces a density through the continuity equation; hence, a violation of this relationship of any kind would not respect local electron number conservation.
 To remedy this defect, Mermin instead proposed an alternate approach where, provided one can define a local chemical potential $\mu(\vec q,\omega)$, it is possible to use the continuity equation to relate the impurity response to the analytically continued density response function as
\begin{equation}
\chi^{imp}(\vec q, \omega) = \frac{\chi(\vec q, \omega + \frac{i}{\tau})}{1+ (1- i \omega \tau)^{-1}\left( \frac{\chi(\vec q, \omega + \frac{i}{\tau})}{\chi(\vec q, 0)}-1 \right)}.
\label{Eq:ImpurityResponse}
\end{equation}   
It can be clearly seen that, in general, $\chi^{imp}(\vec q, \omega ) \neq \chi(\vec q, \omega + \frac{i}{\tau})$. The two of them become the same only when the density response without impurities is energy independent.  \\ \newline
\textit{Electron gas:} To lead our discussion toward a flat band response, let us quickly recall the linear response behavior of a $d$-dimensional electron gas in the presence of scalar impurities. By substituting the $d-$dimensional Lindhard function, $\chi_0(\vec q, \omega)$, in place of the density response into Eq.~\ref{Eq:ImpurityResponse} and making the replacement $\chi_0(\vec q, \omega + i\eta) \rightarrow \chi_0(\vec q, \omega + \frac{i}{\tau})$, we obtain in the diffusive limit ( $\omega \tau\ll 1$ and $qv_F \tau \ll 1$ where $v_F$ is the Fermi velocity and $q$ is the magnitude of $\vec q$)~\cite{rammer2004quantum}
\begin{equation}
\chi^{imp}_0(\vec q, \omega) \simeq \frac{-N_0 D q^2}{ - i \omega + D q^2}.
\label{eq:DiffusionResponse}
\end{equation}
Here $D$ is the diffusion constant given by $D = \frac{v_F^2 \tau}{d}$, and $N_0$ is the density of states at the Fermi level. An important feature of the form of the impurity response appearing in Eq.~\ref{eq:DiffusionResponse} is the presence of a diffusion pole at $\omega = -i D q^2$. The diffusion pole results in a strong enhancement of the density fluctuation spectrum at low energies and, depending on the spatial dimension, plays a crucial role in drastically modifying the quasiparticle scattering rates. Importantly, the form of the diffusion pole appearing the denominator of Eq.~\ref{eq:DiffusionResponse} indicates that the electron density, $n(\vec r, t)$, relaxes according to the diffusion equation given by
\begin{equation}
\frac{\partial n(\vec r, t)}{\partial t} = D \nabla^2 n(\vec r, t),
\label{eq:DiffusionEquation}
\end{equation}
and the average mean square electron displacement scales as $\langle \vec r~^2(t) \rangle \sim 2 d D t$. \newline \\ 
\textit{Flat band with no interactions:} To begin our analysis for the impurity diffusion in a free flat band, it is useful to  provide a precise definition of what we mean by a flat band at the very outset. At an operational level, we define a flat band as a system satisfying two conditions:\\ \newline (i) the bandwidth/dispersion is the smallest energy scale in the problem, i.e., if $\epsilon(\vec k)$ is the energy dispersion, $\epsilon(\vec k) - \epsilon(\vec k + \vec q) \equiv \gamma \ll \omega, \tau^{-1}, T$ and, \\ \newline (ii) $\gamma\rightarrow 0$ and $\Lambda^d \equiv a^{-d} \rightarrow \infty$ such that the product $\gamma \Lambda^d $ goes to a constant ($a$ is the lattice spacing).\\ \newline From now on, we will choose $\gamma$ to be a small constant that tends to zero. It is even possible, in principle, to supplement this constant with momentum dependence.  Our results will, however, remain unchanged as long as its width goes to zero by satisfying the two above conditions. Note that as a consequence of the conditions stated above, the zero temperature dependence appearing in the sections below are obtained by \textit{first} taking $\frac{\epsilon(\vec k)}{T}\rightarrow0$ \textit{and then} $\frac{T}{E}\rightarrow 0$, where $E$ could be any other remaining energy scale in the problem. Interchanging this order of limits will, in general, yield results not relevant to a flat band. With these points in mind and noting that the system is set at half filling by choosing the chemical potential $\mu$ to zero, it is easy to write the non-interacting flat band Lindhard function as  (in the limit $\gamma\rightarrow0$)
\begin{equation}
\chi_0\left(\omega+ \frac{i}{\tau}\right)_{FB} = -\frac{R}{t_0^2}.
\label{eq:FreeFlatBand}
\end{equation}
Here we have defined the \textit{flat band constant} $R = 2 \gamma \Lambda^d \tau^2$ which takes a finite non-zero value in the $\gamma \rightarrow 0$ limit and characterizes the flat band, and $t_0 = t + i$ where $t\equiv \omega \tau$. It is worth noticing that,  unlike the case of the $d-$dimensional electron gas, both the flat band Lindhard function as well as its resulting impurity response function (that appears below) is independent of the momentum transfer $\vec q$. This is not unexpected given that, by definition, a flat band has no spatial dynamics. To obtain the impurity response for the flat band, we also need the flat band Lindhard function at zero frequency. This quantity can be seen to diverge as $\chi_0(0)_{FB} = \frac{R}{\gamma^2}$ as $\gamma\rightarrow 0$ due to the chemical potential being set to zero. Substituting for $\chi_0\left(\omega+ \frac{i}{\tau}\right)_{FB}$ and $\chi_0(0)_{FB}$ into Eq~\ref{Eq:ImpurityResponse}, we obtain in the diffusion limit ($\omega\tau\ll 1$)
\begin{equation}
\chi_0^{imp}(\omega)_{FB} \simeq \frac{R}{-i t} = \frac{R'}{-i \omega},
\label{eq:FreeFlatBandImpurity}
\end{equation}
where $R' \equiv R/\tau$. The real part of $\chi_0^{imp}(\omega)_{FB}$ (of order $R$) has not been included in Eq~\ref{eq:FreeFlatBandImpurity} as it is of order $\omega\tau$ smaller than the imaginary part. Through a comparison of the structure of the poles in Eqs~\ref{eq:FreeFlatBandImpurity} and \ref{eq:DiffusionResponse}, one can conclude that the corresponding electron density, $n(\vec r, t)$, does not relax according to the diffusion equation Eq~\ref{eq:DiffusionEquation}, entirely due to the lack of spatial dynamics. Therefore, in the strictest sense, there is no electron diffusion in a flat band. However, Eq.~\ref{eq:FreeFlatBandImpurity} is still a meaningful quantity as the weight factor $R'$ (or $R$) plays the analogous role of the diffusion constant $D$ times the density of states at the Fermi level, and can therefore be extracted experimentally. $R'$ also plays a central role in defining the characteristics of the electron gas and its properties in the presence of long range interactions will be further explored in the next section. A significant feature of the flat band impurity response (in Eq.~\ref{eq:FreeFlatBandImpurity}) is that the `diffusion' pole shifts to zero energy as opposed to a non-zero momentum dependent value in a $d-$dimensional electron gas. This shift to zero energy is expected to result in a serious toll on the quasiparticle lifetime near the Fermi surface, which is already shortened considerably in a $d-$dimensional electron gas in the presence of impurities. A more quantitative analysis of the quasiparticle lifetime in a dirty flat band is beyond the scope of this paper and will be the focus of future work.  \newline \\

The susceptibility bubble for a flat band in the presence of long range interactions is given by (FBLR denotes flat band, long range)
\begin{equation}
\chi_0(\vec q, iq_n)_{FBLR} = \frac{1}{\beta V} \sum_{ik_n \vec k \sigma} G_0(\vec k,ik_n)_U  G_0(\vec k + \vec q,ik_n + i q_n)_U.
\end{equation}
This expression can be evaluated in the flat band limit and found to be momentum independent (just as in the case of the free flat band) and will be denoted as $\chi_0(\omega)_{FBLR}$ after analytic continuation. The expression for $\chi_0(\omega)_{FBLR}$ can be substituted into Eq.~\ref{Eq:ImpurityResponse} to yield an impurity response (see Appendix A) of  the form
\begin{equation}
\chi_0^{imp}(\omega)_{FBLR} \simeq -\frac{R t_0}{t} \left[ \frac{\kappa_0(g)}{t_0^2} + \frac{\kappa_+(g)}{t_+^2} + \frac{\kappa_-(g)}{t_-^2} \right].
\label{eq:ImpurityFBLR}
\end{equation} 
Here we have defined the dimensionless variables $t_{\pm} \equiv t \pm u + i$, where $u\equiv U \tau$ and  $\kappa_0,\kappa_{\pm}$ are functions of $g \equiv g(U) = \frac{2}{3+ e^{-\beta U}}$, and $t$ and $R$ have been previously defined. The functions $\kappa_0,\kappa_{\pm}$ are given by
\begin{eqnarray}
\kappa_0(g) &=& \left(g^2 +\frac{2(1-g)^2}{e^{\beta U} + 1}\right)\\
\kappa_+(g) &=& \frac{2g(1-g)}{e^{\beta U} + 1}\\
\kappa_-(g) &=& g(1-g).
\end{eqnarray}
There are several aspects of $\chi_0^{imp}(\omega)_{FBLR}$ appearing in Eq.~\ref{eq:ImpurityFBLR} that are worthy of mention. Firstly, the general form of Eq.~\ref{eq:ImpurityFBLR} bears some similarities to Eq.~\ref{eq:FreeFlatBandImpurity}. While the first term is just the free flat band result times a $g$ (and hence temperature) dependent 'weight' factor, $\kappa_0(g)$, the rest of the two terms are modified by the presence of a non-zero interaction strength $U$. These two terms also appear with their own weight factors $\kappa_{\pm}(g)$. Thus, the impurity response of a flat band system, in the presence of a constant long range interactions, is distributed between pure flat band and interaction-modified flat band terms with their respective weights. For $U=0$ we have $\kappa_0 = \frac{1}{2}$ and $\kappa_{\pm} = \frac{1}{4}$ and, as must be expected, $\chi_0^{imp}(\omega)_{FBLR}$ reduces to $\chi_0^{imp}(\omega)_{FB}$. Secondly, it is useful to study the expression for $\chi_0^{imp}(\omega)_{FBLR}$ in different limits. For a repulsive, finite long range interaction ($U>0$) and $T\rightarrow0$, we have $\kappa_0\rightarrow \frac{4}{9}$, $\kappa_+ \rightarrow 0$ and $\kappa_- \rightarrow \frac{2}{9}$. Thus, the maximum fraction of the response comes from the free flat band while there is no contribution from the $\kappa_+$ term. If we extend the repulsive interaction to $U\rightarrow \infty$, then the contribution from the $\kappa_-$ term is negligible, and we are left with $\chi_0^{imp}(\omega)_{FBLR}\simeq \frac{4}{9}\chi_0^{imp}(\omega)_{FB} < \chi_0^{imp}(\omega)_{FB} $. Thus, one can define an effective flat band constant $\tilde{R}$ which is reduced by a factor of $\frac{4}{9}$ compared to the free flat band value. Similarly when $T\rightarrow0$ for a repulsive interaction with $U\ll\omega$, we have $\chi_0^{imp}(\omega)_{FBLR}\simeq \frac{2}{3}\chi_0^{imp}(\omega)_{FB} < \chi_0^{imp}(\omega)_{FB} $. It is also possible to obtain an enhanced effective flat band constant by choosing an attractive interaction $U$. For example, in the limit when $U<0$ and $T\rightarrow0$, we have $\kappa_0 \rightarrow 2$ and $\kappa_{\pm}\rightarrow0$. This results in $\chi_0^{imp}(\omega)_{FBLR}\simeq 2\chi_0^{imp}(\omega)_{FB} > \chi_0^{imp}(\omega)_{FB}$, where the effective flat band constant is twice the free flat band value. This state of affairs obtains because choosing $U<0$ amounts to each $\vec k$ point being pairwise occupied and hence doubling the effective flat band constant and the impurity response. In contrast, when $U>0$ and large, single occupancy of each momentum point is energetically more favorable, and hence the response is lower than the free flat band value. \\ \newline
\textit{Momentum dependent interaction $U(\vec k)$:} In the previous section, we assumed the interaction $U$ to be a constant. It would be more meaningful to explore the effect of a momentum-dependent interaction as in the case of a Coulomb interaction. To this end, we modify our interaction to be of the more generalized Yukawa form $U(\vec k) = U + \frac{\alpha}{\lambda^2 + k^2}$  (this amounts to a Coulomb interaction when $U=\lambda =0$), so that we recover Eq~\ref{eq:ImpurityFBLR} for $\chi_0^{imp}(\omega)_{FBLR}$ when $\alpha=0$. Formally, such a generalization from a constant $U$ to $U(\vec k)$ is fairly straightforward$-$ the variable $U$ appearing in the momentum space representation of the Hamiltonian in Eq.~\ref{eq:kSpaceHK} and Green function in Eq.~\ref{eq:GreensFunction} has to be replaced by $U(\vec k)$. However, the ensuing momentum integrals are fairly convoluted and can only be solved in certain limits and simplifying assumptions. Once we understand how the integrals behave in these limits, we can gain insight into what they look like for other parallel cases. For analytical tractability, we will assume that $U$ is repulsive so that $U(\vec k)$ is always positive. In this limit, the momentum integrals can be solved exactly as $T\rightarrow0$ in all three dimensions (see Appendix B) and the response (which we donote as $\chi_0(\omega)_{FBY}$, where $FBY$ stands for Flat Band Yukawa) in the presence of the interaction $U(\vec k)$ becomes
\begin{widetext}
\[
\chi_0(\omega)_{FBY}\simeq
\begin{cases}
\hphantom{-}\chi_0(\omega)_{FBLR} + lim_{\gamma\rightarrow0}\left[ \frac{\pi \alpha}{z'^2} \sqrt{\lambda^2 + \frac{\alpha}{z'}} - \frac{\pi \alpha}{z^2} \sqrt{\lambda^2 - \frac{\alpha}{z}} \right]+ \mathscr{O}(\frac{1}{\Lambda})&\text{$d=3$},\\[2ex]
\hphantom{-}\chi_0(\omega)_{FBLR} +  lim_{\gamma\rightarrow0}\left[ \alpha~Log\left(\Lambda\right)\left[\frac{1}{z^2} - \frac{1}{z'^2}\right]\right] +\mathscr{O}(\frac{1}{\Lambda}) &\text{$d=2$},\\[2ex]
\hphantom{-}\chi_0(\omega)_{FBLR} +  lim_{\gamma\rightarrow0}\left[\frac{\pi \alpha}{z^2 \sqrt{\lambda^2 - \frac{\alpha}{z}}} - \frac{\pi \alpha}{z'^2 \sqrt{\lambda^2 + \frac{\alpha}{z'}}}\right] + \mathscr{O}(\frac{1}{\Lambda}) &\text{$d=1$}.
\end{cases}
\]
\end{widetext}
Here we have defined $z = -\omega - \frac{i}{\tau} + \gamma + U$, $z' = \omega + \frac{i}{\tau} +\gamma - U$, and $\Lambda$ is the large momentum cutoff. To derive the above expressions, we have only kept leading order contributions in $\Lambda$. For each dimension $d$ except $d=2$, the largest contributing terms are of $\mathscr{O}(\Lambda^d)$ and the next highest order is $\mathscr{O}(1)$. In the case of $d=2$, the next highest order is of $\mathscr{O}(Log(\Lambda))$. However, in each dimension, the $\mathscr{O}(\Lambda^d)$ terms recombine to give $\chi_0(\omega)_{FBLR}$, which is $\mathscr{O}(1)$ because we have assumed that $\Lambda^d \gamma$ is a constant of order unity.  We must therefore keep the next highest order terms in each case. However, in the limit $\gamma \rightarrow 0$, we have $z' = -z$, and according to our definition of a flat band, all the terms proportional to $\alpha$ vanish (see the formula for $\chi_0(\omega)_{FBY}$ above). \textit{Thus we can conclude that a momentum-dependent interaction of the form $\frac{\alpha}{\lambda^2 + k^2}$ has no effect on the impurity response in the flat band limit}. Although our conclusion is derived for the case when $U$ is repulsive and $T\rightarrow0$, it is easy to see why it holds for other cases as well. The key reason why an interaction of the Yukawa or Coulomb form does not affect the flat band response function (with or without impurities) is that only electrons near the Brillouin zone edges contribute to the response. This can be seen from from our assumption that $\Lambda^d \gamma$ is a constant; hence, electrons not at the boundary do not contribute because $\gamma$ goes to zero faster than the enclosed zone volume at finite $\vec k$. Thus, given that only the edge electrons contribute in a flat band, in the limit that $\Lambda^d\rightarrow \infty$, $\frac{\alpha}{\lambda^2 + k^2}$ has no effect as it goes to zero at the Brillouin zone edges. Therefore a potential of the form of $\frac{\alpha}{\lambda^2 + k^2}$ does not change a flat band response with or without impurities for either attractive or repulsive interactions, even at non-zero temperatures. However, when $U(\vec k)$ is a constant (as in the previous section), the potential has a finite value even at the zone edges and, as a consequence, has a non-trivial effect on the response properties.  The irrelevance of a Yukawa or Coulomb-like term is exclusively a property of the flat band and does not hold in the case of a dispersive band due to the fact that the $\mathscr{O}(1)$ $\alpha$ term contributions are, in general, finite and non-zero. \\ \newline
\begin{figure}[h!]
\includegraphics[width=3.5in,height=2 in]{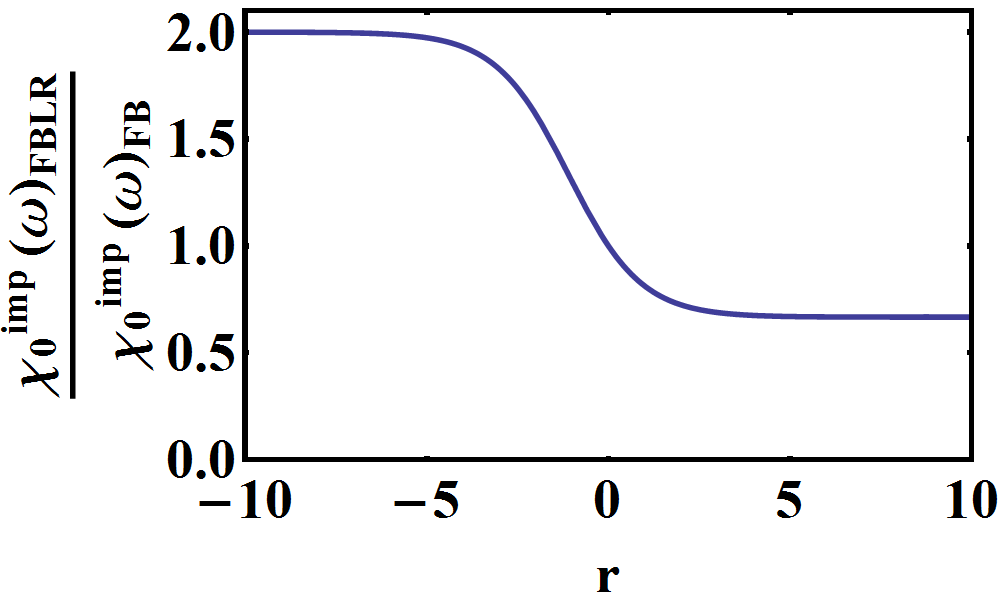}
\caption{Plot of $\chi_0^{imp}(\omega)_{FBLR}$ normalized with the free flat band impurity response (and hence independent of $\omega$) as  function of the dimensionless parameter $r= \beta U$. Positive (negative) $r$ corresponds to repulsive (attractive) long range interaction.  }\label{Temperature}
\end{figure}
\textit{Temperature dependence:} For $d-$dimensional electrons, in the absence of any interactions, the band width sets the only energy scale in the problem. It is with respect to this scale that electrons can be thermally excited into higher energy states away from the Fermi level leading to temperature-dependent response functions. However, in the case of a flat band, where all the states have an equal probability of occupation, there is no notion of ordering of states energetically. Hence, the free flat band response with or without impurities is independent of temperature as reflected in Eqs.~\ref{eq:FreeFlatBand} and \ref{eq:FreeFlatBandImpurity}.  An equivalent statement is that in the extremely high temperature limit ($T$ much larger than the band width), the temperature is already large enough so that all momentum states are energetically accessible and, therefore, any smaller changes in temperatures will not have any effect on occupation number dependent observable properties. In the presence of long range interactions of the type appearing in Eq.~\ref{eq:kSpaceHK}, however, $U$ sets the only scale in the problem. It is, therefore, reasonable to expect that observable quantities depend on temperature through the dimensionless parameter $\beta U \equiv r$; Eq.~\ref{eq:ImpurityFBLR} reflects this expectation. A plot of the $\chi_0^{imp}(\omega)_{FBLR}$ (normalized with the free flat band impurity response, and hence independent of $\omega$) as a function of $r$, for $U$ much smaller than $\omega$, appears in Fig~\ref{Temperature}. There is a significant temperature dependence of $\chi_0^{imp}(\omega)_{FBLR}$ ($\propto \beta$) only for small $r$ and is featureless asymptotically. This behavior can be understood from the form of the function $g(U)$ which determines the fraction of the response  that is split between the free flat band and the part dominated by long-range interactions. \\ \newline
\textit{Conclusions:} To conclude, we studied the impurity response of a flat band in the diffusion limit with and without long range interactions of the type proposed in Ref.~\cite{hk1992}. Starting from the non-interacting case, we argued that the system does not 'diffuse' in the traditional sense of a $d-$dimensional electron gas, but found it useful to define the notion of a \textit{flat band constant} that takes a non-zero value inspite of the fact that the Fermi velocity of a dispersionless band is zero. \textcolor{black}{This constant appears as the residue of an impurity pole that has been shifted to zero energy due to the lack of spatial dynamics and momentum independence of the impurity response}.  In the presence of long range interactions, we saw that the flat band constant could be effectively enhanced, suppressed or unaffected depending on whether the interaction is a constant and attractive, a constant and repulsive, or momentum dependent of the Yukawa/Coulomb form respectively.  \textcolor{black}{The impurity pole at zero energy is unaffected for a constant attractive interaction at zero temperature, for an infinite repulsive interaction, or for an interaction of the Yukawa/Coulomb form. This shows that scale invariance is robust to interactions of these types and there is no energy scale which develops that permits a reasonable definition of a charge diffusion constant}. Finally, we argued the temperature independence of the impurity response for the case of the dirty non-interacting flat band. In the interacting case, the calculated response is $\propto \beta$ for small $\beta U$ and saturates to a constant for larger values. Looking ahead, it would be of interest to further explore the effect of the shift of the diffusion pole to zero energy in the exact scaling form of the quasiparticle life time in dirty flat band systems.  \\ \newline
\textit{Acknowledgments:} We acknowledge support from Center for Emergent
Superconductivity, a DOE Energy Frontier Research Center, Grant No. DE-AC0298CH1088.  We also thank the NSF DMR-1461952 for partial funding of this project.
\bibliographystyle{apsrev4-1}%Choose a bibliograhpic style
\bibliography{FlatBand}
\onecolumngrid
\newpage
\section{APPENDIX A}
The form of the Green function for the model proposed in ref.~\cite{hk1992} has been derived in ref.~\cite{hk1996} which we simply re-state here for completeness. Given a quasiparticle dispersion $\xi(\vec k)$, the Green function is given as
\begin{equation}
G_0(\vec k,i\omega_n)_U = \left(\frac{g(\vec k, U)}{i \omega_n - \xi(\vec k)} +  \frac{1- g(\vec k, U)}{i\omega_n - \left(\xi(\vec k) + U\right)}\right),
\end{equation}
where the function $g(\vec k, U)$ is defined by
\begin{equation}
g(\vec k, U) = \frac{1+ e^{-\beta \xi(\vec k)}}{1+ 2 e^{-\beta \xi(\vec k)} + e^{-\beta\left(2\xi(\vec k) + U\right)}},
\end{equation}
with $\beta$ being the inverse temperature, $i\omega_n$ is the fermionic Matsubara frequency, and $U$ is the constant long range interaction. The density-density correlation function for such a Green function is given (see, for example, Bruus-Flensberg(2004)) by the generalized pair susceptibility bubble 
\begin{eqnarray}
\chi_0(\vec q, i q_n) &=& \frac{1}{\beta V} \sum_{i k_n}\sum_{\vec k \sigma} G_0(\vec k,i\omega_n)_U G_0(\vec k + \vec q,i\omega_n +  iq_n)_U \\ \nonumber
&=& \frac{1}{\beta V} \sum_{i k_n}\sum_{\vec k \sigma} \Biggl[ \frac{g(\vec k, U) g(\vec k + \vec q, U)}{\left(i k_n - \xi(\vec k)\right)\left( i k_n + i q_n - \xi(\vec k+ \vec q ) \right)} + \frac{\left(1-g(\vec k, U)\right)\left(1- g(\vec k + \vec q, U)\right)}{\left(i k_n - \xi(\vec k) -U\right)\left( i k_n + i q_n - \xi(\vec k+ \vec q) - U \right)} \\ 
&+& \frac{g(\vec k, U) \left( 1- g(\vec k + \vec q, U)\right)}{\left(i k_n - \xi(\vec k)\right)\left( i k_n + i q_n - \xi(\vec k+ \vec q )  - U\right)} + \frac{\left(1- g(\vec k, U)\right) g(\vec k + \vec q, U)}{\left(i k_n - \xi(\vec k) - U\right)\left( i k_n + i q_n - \xi(\vec k+ \vec q ) \right)} \Biggr].
\end{eqnarray}
Performing the Matsubara sums and shifting the momentum from $\vec k \rightarrow -\vec k - \vec q$ in the second of each resulting Fermi distribution, we obtain
\begin{eqnarray}
\chi_0(\vec q, \omega + \frac{i}{\tau}) &=& \frac{1}{V} \sum_{\vec k \sigma} \Biggl[g(\vec k, U) g(\vec k + \vec q, U) \Biggl(\frac{n_{\vec k}}{\omega + \frac{i}{\tau} + \xi(\vec k) - \xi(\vec k + \vec q)} + \frac{n_{\vec k}}{-\omega - \frac{i}{\tau} + \xi(\vec k) - \xi(\vec k + \vec q)} \Biggr) \\ \nonumber
&& \!\!\!\!\!\!\!\!\!\!\!\!\!\!\!\!\!\!\!\!\!\!\!\!\!\!\!\!\!\!\!\!\!\!\!\!\!\!\!\!\!\!\!\!\! +  \left(1- g(\vec k, U)\right) \left(1-  g(\vec k + \vec q, U) \right) \Biggl(\frac{n_{\vec k U}}{\omega + \frac{i}{\tau} + \xi(\vec k)_U - \xi(\vec k + \vec q)_U} + \frac{n_{\vec k U}}{-\omega - \frac{i}{\tau} + \xi(\vec k)_U - \xi(\vec k + \vec q)_U} \Biggr) \\ \nonumber
&& \!\!\!\!\!\!\!\!\!\!\!\!\!\!\!\!\!\!\!\!\!\!\!\!\!\!\!\!\!\!\!\!\!\!\!\!\!\!\!\!\!\!\!\!\! +  g(\vec k, U) \left(1-  g(\vec k + \vec q, U)\right) \Biggl(\frac{n_{\vec k}}{ \omega + \frac{i}{\tau} + \xi(\vec k) - \xi(\vec k + \vec q)_U} \Biggr) +  g(\vec k + \vec q, U) \left(1-  g(\vec k, U) \right) \Biggl(\frac{n_{\vec k U}}{ -\omega - \frac{i}{\tau} + \xi(\vec k) - \xi(\vec k + \vec q)_U} \Biggr)  \\ \nonumber
&& \!\!\!\!\!\!\!\!\!\!\!\!\!\!\!\!\!\!\!\!\!\!\!\!\!\!\!\!\!\!\!\!\!\!\!\!\!\!\!\!\!\!\!\!\! +  g(\vec k + \vec q, U) \left(1-  g(\vec k, U)\right) \Biggl(\frac{n_{\vec k U}}{ \omega + \frac{i}{\tau} + \xi(\vec k)_U - \xi(\vec k + \vec q)} \Biggr) +  g(\vec k, U) \left(1-  g(\vec k + \vec q, U) \right) \Biggl(\frac{n_{\vec k }}{ -\omega - \frac{i}{\tau} + \xi(\vec k)_U - \xi(\vec k + \vec q)} \Biggr) \Biggr],
\end{eqnarray}
where we have defined $n_{\vec k} \equiv n_F\left(\xi(\vec k)\right)$, $n_{\vec k U} = n_F\left(\xi_{\vec k} + U\right)$, $\xi(\vec k )_U = \xi(\vec k) + U$, and $n_F(x)$ is the Fermi function. Substituting for $\xi(\vec k) - \xi(\vec k + \vec q) = \gamma$, using the definition of the flat band, and with the help of Eq~\ref{Eq:ImpurityResponse} appearing in the main text, we obtain the flat band response in the presence of impurties and in the absence of interactions as
\begin{equation}
\chi_0^{imp}(\omega)_{FBLR} \simeq -\frac{R t_0}{t} \left[ \frac{\kappa_0(g)}{t_0^2} + \frac{\kappa_+(g)}{t_+^2} + \frac{\kappa_-(g)}{t_-^2} \right].
\label{eq:ImpurityFBLR}
\end{equation} 
Here we have defined the dimensionless variables $t_0 = t + i$, $t_{\pm} \equiv t \pm u + i$, where $t \equiv \omega \tau$, $u\equiv U \tau$ and  $\kappa_0,\kappa_{\pm}$ are functions of $g \equiv g(U) = \frac{2}{3+ e^{-\beta U}}$. The flat band constant $R \equiv2 \gamma \Lambda^d \tau^2$. In deriving the above expression for $\chi_0^{imp}(\omega)_{FBLR}$, we have used the fact that $\xi(\vec k)$ is equal to zero in the Fermi function ($n_{\vec k}$) and hence every $\vec k$ point has a probability of being occupied as $\frac{1}{2}$. Thus the occupied density of electrons translates to a momentum space volume $\Lambda^d$. Similarly, it is easily seen that $\frac{1}{V} \sum_{\vec k \sigma} n_{\vec k U} = \frac{2 \Lambda^d}{e^{\beta U} + 1}$.
\section{APPENDIX B}
As briefly mentioned in the main text, for analytical tractability, we will assume repulsive interactions and work in the limit of zero temperature. In this limit, we can ignore the terms proportional to $n_{\vec k U}$ to the lowest order as they are exponentially smaller than the terms proportional to $n_{\vec k}$; additionally, the function $g(\vec k, U)$ goes to a constant equal to $\frac{2}{3}\equiv g$. With these assumptions, we can write the response function as
\begin{equation}
\chi_0\left(\vec q, \omega + \frac{i}{\tau}\right) = \frac{1}{V} \sum_{\vec k} \Biggl[ g^2 \left( \frac{1}{\omega + \frac{i}{\tau} + \gamma} + \frac{1}{-\omega - \frac{i}{\tau} + \gamma}\right) + g(1-g) \left( \frac{1}{\omega + \frac{i}{\tau} + \gamma - U(\vec k + \vec q)} + \frac{1}{-\omega - \frac{i}{\tau} + \gamma + U(\vec k) }\right)\Biggr],
\end{equation}
where $U(\vec k) = U +  \frac{\alpha}{\lambda^2 + k^2}$. The first term proportional to $g^2$ is independent of $\vec k$ and the momentum sum simply gives the non-interacting flat band response modified by a factor of $g^2= \frac{4}{9}$. The effect of the interactions appears in the second term proportional to $g(1-g)$ and has a momentum dependence due to $U(\vec k)$. Hence the resulting momentum integrals need to be performed in all three dimensions. \\ \newline
\underline{\textit{d=3 case:}} Consider the integral
\begin{equation}
I_1(3D) = \frac{g(1-g)}{\left(2 \pi \right)^3} \int_0^{\Lambda} \int_{\Omega_{d=3}} \left[ \frac{dk k^2 d\Omega_{d=3}}{\omega + \frac{i}{\tau} + \gamma - \left(U + \frac{\alpha}{ \left(\vec k + \vec q\right)^2 + \lambda^2}\right)}\right].
\end{equation}
Defining $z\equiv \left( \omega + \frac{i}{\tau} + \gamma -U  \right)$, performing the azimuthal $\phi$ integral, and changing variables $t =  cos\theta$, we obtain
\begin{equation}
I_1(3D) = \frac{g(1-g)}{4 \pi^2} \int_0^{\Lambda} \int_{-1}^{1} \frac{dk k^2 dt \left(X+ Yt \right)}{z\left( X + Y t\right) - \alpha},
\end{equation}
 where we have defined $X\equiv k^2 + q^2 + \lambda^2$, $Y =  2 kq$. The $t$ integral can easily be performed to give
\begin{equation}
I_1(3D) = \frac{g(1-g)}{4 \pi^2} \int_0^{\Lambda} k^2 dk \left[ \frac{2}{z} + \frac{\alpha}{2 k q z^2} Log\left(\frac{z(k+q)^2 + \lambda^2 z - \alpha}{z(k-q)^2 + \lambda^2 z - \alpha} \right) \right].
\end{equation}
The radial momentum integral of the first term is simply $\frac{2 \Lambda^3}{3 z}$. The radial momentum integrals of the second term are given as
\begin{equation}
\int_0^{\Lambda} k~dk Log\left( \mu k^2 \pm \nu k + \delta \right) = \biggl[ I_+(k) -  I_-(k) \biggr]_0^{\Lambda},
\end{equation}
where,
\begin{equation}
I_{\pm}(k) = \frac{1}{4 \mu^2} \left[2 k \mu (- k \mu \pm \nu) \mp 2 \nu \sqrt{4 \delta \mu - \nu^2}~Arctan\left( \frac{2 k \mu \pm \nu}{\sqrt{4 \delta \mu -  \nu^2}}\right) + \left( 2 \mu (\delta + k^2 \mu) - \nu^2\right) Log\left( \delta + k^2 \mu \pm k \nu\right)\right]
\end{equation}
and $\mu\equiv z$, $\nu\equiv 2 q z$ and $\delta \equiv z\left(q^2 + \lambda^2 \right) - \alpha$. Substituting for $I_{\pm}(k)$ into the integral, taking the limit of $\Lambda\rightarrow\infty$ and keeping only the most divergent terms in $\Lambda$ we obtain
\begin{equation}
I_1(3D) \simeq \frac{g(1-g)}{4 \pi^2} \left( \frac{2 \Lambda^3}{3 z} + \frac{2 \Lambda \alpha}{z^2} \right) + \mathscr{O}(\Lambda^0).
\end{equation}
Similarly, the integral 
\begin{equation}
I_2(3D) = \frac{g(1-g)}{\left(2 \pi \right)^3} \int_0^{\Lambda} \int_{\Omega_{d=3}} \left[ \frac{dk k^2 d\Omega_{d=3}}{-\omega - \frac{i}{\tau} + \gamma + \left(U + \frac{\alpha}{ \vec k^2 + \lambda^2}\right)}\right]\simeq  \frac{g(1-g)}{4 \pi^2} \left( \frac{2 \Lambda^3}{3 z'} - \frac{2 \Lambda \alpha}{z'^2} \right) + \mathscr{O}(\Lambda^0),
\end{equation}
where we have defined $z' \equiv -\omega -\frac{i}{\tau} + \gamma + U$. Note that $I_1(3D)$ and $I_2(3D)$ differ by a sign in the second term proportional to $\alpha$ and $z$ is replaced with $z'$. Thus, combining $I_1(3D)$ and $I_2(3D)$, we see that the terms proportional to $\Lambda$ cancel out, whereas terms proportional to $\Lambda^3$ combine with $\gamma$ to give a constant of $\mathscr{O}(1)$. Therefore, we need to include other $\mathscr{O}(1)$ contributions which were left out in $I_1(3D)$ and $I_2(3D)$. These $\mathscr{O}(1)$ contributions can be evaluated to be $\frac{\pi \alpha}{z'^2}\sqrt{\lambda^2 + \frac{\alpha}{z'}} - \frac{\pi \alpha}{z^2}\sqrt{\lambda^2 - \frac{\alpha}{z}}$, which is the expression that appears in the main text. Note that there is an additional prefactor of $\frac{1}{6\pi^2}$ that appears in these expressions compared to free flat band case. This is only because of the difference in the normalization used while working in spherical co-ordinates, and can be trivially absorbed into the defintion of the flat band constant $R$.\\ \newline
\underline{\textit{d=2 case:}} As the remaining two cases can be derived similar to the $d=3$ case, we will only give a sketch of what the integrals look like. In the $d=2$ case we consider the integral
\begin{equation}
I_1(2D) = \frac{g(1-g)}{4 \pi^2}\int_0^{\Lambda} \int_0^{2\pi}\frac{k~dk~d\phi}{z - \frac{\alpha}{\left(\vec k + \vec q\right)^2 + \lambda^2}} =  \frac{g(1-g)}{4 \pi^2} \int_0^{\Lambda}\int_0^{2\pi} \frac{k~dk~d\phi ~(X + Y cos\phi)}{z(X+ Y cos\phi) - \alpha},
\end{equation}
where again $z\equiv \left( \omega + \frac{i}{\tau} + \gamma -U  \right)$, and $X\equiv k^2 + q^2 +\lambda^2$, $Y \equiv 2 k q$. The azimuthal $\phi$ integral can be performed to give
\begin{equation}
I_1(2D) = \frac{g(1-g)}{4\pi^2} \int_0^{\Lambda} k~dk\left[ \frac{2\pi}{z} + \frac{\alpha}{z \sqrt{z(k+q)^2 + z \lambda^2 -\alpha} \sqrt{z(k-q)^2 + z \lambda^2 - \alpha}}\right].
\end{equation}
In the large $\Lambda$ limit, the remaining radial $k$ integral yields 
\begin{equation}
I_1(2D)\simeq \frac{g(1-g)}{4 \pi^2} \left[\frac{\pi \Lambda^2}{z} + \frac{\alpha}{z^2}Log(\Lambda)\right]
\end{equation}
where we have collected the two largest contributions in $\Lambda$. Similarly, we have
\begin{equation}
I_2(2D)= \frac{g(1-g)}{4 \pi^2}\int_0^{\Lambda} \int_0^{2\pi}\frac{k~dk~d\phi}{z' +\frac{\alpha}{\vec k^2 + \lambda^2}}\simeq \frac{g(1-g)}{4 \pi^2} \left[\frac{\pi \Lambda^2}{z'} - \frac{\alpha}{z'^2}Log(\Lambda)\right].
\end{equation}
Combining $I_1(2D)$ and $I_2(2D)$, we see that the terms proportional to $\Lambda$ are multiplied by $\gamma$ and are just of order unity. And the remaining terms proportional to $\alpha Log(\Lambda)$  just obtain $\alpha Log(\Lambda) \left[\frac{1}{z^2} - \frac{1}{z'^2}\right]$ which is the expression that appears in the main text.\\ \newline
\underline{\textit{d=1 case:}} The one dimensional case follows along similar lines. Consider the integral
\begin{equation}
I_1(1D) = \int_{-\frac{\Lambda}{2}}^{\frac{\Lambda}{2}} \frac{dk}{z - \frac{\alpha}{(k+q)^2 + \lambda^2}} =  \int_{\frac{-\Lambda}{2}}^{\frac{\Lambda}{2}} \frac{d k \left(k^2 + a k +b \right)}{z\left(k^2 + a k + b \right) - \alpha},
\end{equation}
where we have defined $a=2q$ and $b=q^2 + \lambda^2$. This integral can be performed exactly and is written as
\begin{equation}
I_1(1D) = \left[\frac{k}{z} +  \frac{2 \alpha~Arctan\left(\frac{(a+ 2 k)\sqrt{z}}{\sqrt{-a^2 z + 4 b z - 4 \alpha}}\right)}{z^{3/2}\sqrt{-a^2 z + 4 b z - 4 \alpha}}\right]_{-\frac{\Lambda}{2}}^{\frac{\Lambda}{2}}.
\end{equation}
In the large $\Lambda$ limit, the leading terms are given as
\begin{equation}
I_1(1D) \simeq \left( \frac{\Lambda}{z} + \frac{2 \pi \alpha}{z^{3/2} \sqrt{-a^2 z + 4 b z - 4 \alpha}}\right)+  \mathscr{O}(\frac{1}{\Lambda}).
\end{equation}
Note that we have also included the $\mathscr{O}(1)$ contributions since we learned from the previous two cases that the $\mathscr{O}(\Lambda^d)$ terms combine to give $\mathscr{O}(1)$ terms. On similar lines we can write
\begin{equation}
I_2(1D) = \int_{\frac{-\Lambda}{2}}^{\frac{\Lambda}{2}} \frac{dk}{z' + \frac{\alpha}{k^2 + \lambda^2}} \simeq \left( \frac{\Lambda}{z'} - \frac{2 \pi \alpha}{z'^{3/2} \sqrt{-a^2 z' + 4 b z' + 4 \alpha}}\right)+  \mathscr{O}(\frac{1}{\Lambda}).
\end{equation}
As before, adding the $I_1(1D)$ and $I_2(1D)$ terms, we see that the terms proportional to $\Lambda$ combine with $\gamma$ gives us a term of $\mathscr{O}(1)$. The remainder of the terms yield $2 \pi \alpha \left[\frac{1}{z^2 \sqrt{4 \lambda^2 - \frac{4\alpha}{z}}} - \frac{1}{z'^2 \sqrt{4 \lambda^2 + \frac{4\alpha}{z'}}} \right]$ which is the expression appearing in the main text.
\end{document}